\documentclass[11pt, onecolumn, draftcls, singlespace ]{IEEEtran}
\usepackage{cite}
\usepackage{algorithm2e}
\usepackage{amssymb}
\usepackage{enumerate}
\usepackage{graphicx}
\usepackage{amssymb}
\usepackage{amsbsy}
\usepackage{float}
\usepackage{balance}
\usepackage{url}
\usepackage{array}
\usepackage{varioref}
\usepackage[normalem]{ulem}
\graphicspath{{./figs/}}
\usepackage{color}
\usepackage{empheq}
\usepackage{flushend}
\begin{document}

\title{ Sum Rate Maximizing  Multigroup Multicast Beamforming under Per-antenna Power Constraints  }
 \author{\IEEEauthorblockN{ Dimitrios Christopoulos\IEEEauthorrefmark{1},  Symeon Chatzinotas\IEEEauthorrefmark{1}and Bj\"{o}rn Ottersten\IEEEauthorrefmark{1}}
%
 \IEEEauthorblockA{\IEEEauthorrefmark{1}SnT - securityandtrust.lu,  University of Luxembourg
 \\email: \textbraceleft dimitrios.christopoulos, symeon.chatzinotas, bjorn.ottersten\textbraceright@uni.lu}
}
\maketitle
\begin{abstract}
  A multi-antenna transmitter that conveys independent sets of common data to distinct groups of  users is herein considered, a model known as physical layer multicasting to multiple co-channel groups. In the recently proposed context of per-antenna power constrained multigroup multicasting, the present work focuses on a novel system design that aims at maximizing the total achievable throughput.
Towards increasing the system sum rate,  the available power resources need to be allocated to well conditioned groups of users. A detailed solution to tackle  the elaborate sum rate maximization multigroup multicast problem under per-antenna power constraints is therefore derived. Numerical results are presented to quantify the gains of the proposed algorithm over heuristic solutions. Besides  Rayleigh faded channels, the solution  is also applied to uniform linear array transmitters operating in the far field, where line-of-sight conditions are realized. In this setting, a sensitivity analysis with respect   to the angular separation of co-group users is included. Finally, a simple scenario  providing important intuitions for the sum rate maximizing multigroup multicast solutions is elaborated.
\end{abstract}

\section{Introduction \& Related Work}
 Advanced transmit signal processing techniques are currently  employed to optimize the performance of  multi-antenna transmitters without compromising the complexity of single antenna receivers.
These   beamforming (or equivalently precoding) techniques efficiently manage the co-channel interferences to achieve the targeted service requirements  (Quality of Service--$\mathrm{QoS}$ targets).
The optimal downlink transmission strategy, in the sense of minimizing the total transmit power under guaranteed per  user $\mathrm{QoS}$ constraints,    was given in \cite{Bengtsson2001,bengtsson1999}.
Therein, the  powerful tool of Semi-Definite Relaxation ($\mathrm{SDR}$) reduced the non-convex quadratically constrained quadratic  problem ($\mathrm{QCQP}$)  into a relaxed semi-definite programming  instance by changing the optimization variables and disregarding the unit-rank constraints over the new variable. The relaxed solution  was proven to be optimal.  {In the same direction, the multiuser downlink beamforming problem that aims at maximizing the minimum over all users signal to  interference plus noise ratio $(\mathrm{SINR}$),  was optimally solved in \cite{Schubert2004}. The goal of the later  formulation is to increase the fairness of the system by boosting the  $\mathrm{SINR}$ of the  user that is further away from a targeted performance. Hence,  the problem is commonly referred to as \textit{max--min fair}.

In the  contributions discussed so far, power flexibility amongst the transmit antennas is a    fundamental assumption. Hence, in all the  above optimization problems a  sum power constraint ($\mathrm{SPC}$) at the transmitter is imposed. The more elaborate transmit beamforming problem under per-antenna  power constraints ($\mathrm{PAC}$s) was formulated and solved   in \cite{Yu2007}. The motivation for the $\mathrm{PAC}$s originates from  practical system implementation aspects. The lack of flexibility in sharing energy resources amongst the antennas of the transmitter is usually the case. Individual amplifiers per  antenna are  common practice.  Although flexible amplifiers could be incorporated in multi-antenna transmitters, specific communication systems cannot afford this design. Examples of such systems can be found in satellite communications, where highly complex payloads are restrictive and in distributed antenna systems where the  physical co-location of the transmitting elements is not a requisite.

 In the new generation of multi-antenna communication standards, the adaptation of  the physical layer design to the needs of the higher network  layers can significantly enhance the system capabilities. Physical layer ($\mathrm{PHY}$) multicasting has the potential to efficiently address the nature of traffic demand in these systems and  has become part of the evolution of communication standards. An inherent  consideration of the hitherto presented literature is that  independent  data is addressed to multiple users.      When a symbol is addressed to more than one users,  however,  a more elaborate problem formulation is emanated. In this direction, the  $\mathrm{PHY}$ multicasting problem was proposed, proven NP-hard and accurately approximated by $\mathrm{SDR}$ and Gaussian randomization techniques in \cite{Sidiropoulos2006}.
Following this, a unified framework for physical layer multicasting to multiple interfering groups, where independent sets of common data are transmitted to  multiple interfering groups of users by the multiple antennas, was presented in \cite{Karipidis2008}. Therein, the  $\mathrm{QoS}$ and the \textit{max--min fair} problems were formulated, proven NP-hard and accurately approximated for  the $\mathrm{SPC}$ multicast multigroup case. Extending these works, a consolidated solution for the weighted \textit{max--min fair} multigroup multicast beamforming under $\mathrm{PAC}$s
has been derived in \cite{Christopoulos2014,Christopoulos2014ICC}. To this end, the well established tools of $\mathrm{SDR}$ and Gaussian randomization where combined with bisection to obtain highly accurate and efficient solutions.

The fundamental consideration of  multicasting, that is    a  single transmission addressing a group of users, constrains the system performance according to the worst user.  Therefore, the maximization of the minimum $\mathrm{SINR}$ is the most relevant problem and the fairness criterion is imperative. When advancing to multigroup multicast systems, however, the service levels between different groups can be adjusted towards achieving some other optimization goal. The consideration to maximize the total system sum rate in a multigroup multicast context was initially considered in  \cite{Kaliszan2012} under $\mathrm{SPCs}$. Therein, a heuristic iterative algorithm was developed based on the principle of decoupling the beamforming design and the power allocation problem.
 In the present contribution, the focus is set on maximizing the total  throughput of the multigroup multicast system under $\mathrm{PAC}$s. To this end, the \textit{max sum rate}  ($\mathrm{SR}$) multigroup multicast problem under  $\mathrm{PAC}$s is formulated and solved.

{\textit{Notation}: In the remainder of this paper, bold face lower case and upper case characters denote column vectors  and matrices, respectively. The operators \(\left(\cdot\right)^\text{T}\), \(\left(\cdot\right)^\dag\), $|\cdot|$ and \(|| \cdot ||^2_2 \) correspond to   the transpose, the conjugate transpose,  the absolute value and the  Frobenius norm of matrices and vectors, while $[\cdot]_{ij}  $  denotes the $i, j$-th element of a matrix. $\mathrm{Tr}(\cdot)$  denotes the trace operator over square matrices and $\mathrm{diag(\cdot)}$ denotes a square diagonal matrix with elements that of the input vector. Calligraphic indexed characters denote sets}.

\section{System Model}\label{sec: System Model}
 Herein, the focus is on a multi-user ($\mathrm{MU}$) multiple input single output ($\mathrm{MISO}$) multicast system. Assuming a single transmitter, let   $N_t$ denote the number of transmitting elements  and  $N_{u}$ the  total number of users served.  The input-output analytical expression  will read as $y_{i}= \mathbf h^{\dag}_{i}\mathbf x+n_{i},$
where \(\mathbf h^{\dag}_{i}\) is a \(1 \times N_{t}\) vector composed of the channel coefficients (i.e. channel gains and phases) between the \(i\)-th user and the  \(N_{t}\) antennas of the transmitter, \(\mathbf x\) is the \(N_{t} \times 1\)  vector of the transmitted symbols and  \(n_{i}\) is the independent complex circular symmetric (c.c.s.) independent identically distributed (i.i.d) zero mean  Additive White Gaussian Noise ($\mathrm{AWGN}$),  measured at the \(i\)-th user's receive antenna.

Focusing  in a multigroup multicasting scenario,  let there be a total of $1\leq G \leq N_{u}$ multicast groups with  $\mathcal{I} = \{\mathcal{G}_1, \mathcal{G}_2, \dots  \mathcal{G}_G\}$ the collection of   index sets and $\mathcal{G}_k$ the set of users that belong to the $k$-th multicast group, $k \in \{1\dots G \}$. Each user belongs to only one group, thus $\mathcal{G}_i\cap\mathcal{G}_j=$\O ,$  \forall i,j \in \{1\cdots G\}$. Let $\mathbf w_k \in \mathbb{C}^{N_t \times 1}$ denote the precoding weight vector applied to the transmit antennas to beamform towards the $k$-th group.
By collecting all user channels in one channel matrix, the general linear signal model in vector form reads as
$ \mathbf y = \mathbf {H}\mathbf x + \mathbf n = \mathbf {H}\mathbf {Ws} + \mathbf{ n}$,
where $ \mathbf {y \text{ and }  n } \in \mathcal{\mathbb{C}}^{N_{u}}$, $\mathbf {x} \in \mathbb{C}^{N_{t}}$ and $ \mathbf {H} \in \mathbb{C}^{N_{u} \times N_t}$.  The multigroup multicast scenario imposes  a precoding matrix $ \mathbf {W} \in \mathbb{C}^{N_t  \times G}$ that includes as many precoding vectors (i.e columns) as the number of groups. This is the number of independent symbols transmitted, i.e. $\mathbf {s} \in \mathbb{C}^{G}$.
The assumption of independent information transmitted to different groups implies that the symbol streams $\{s_k\}_{k=1}^G$ are mutually uncorrelated and the total power radiated from the antenna array is equal to
\begin{align}
P_{tot} = \sum_{k=1}^ G \mathbf w_k \mathbf w_k^\dag= \mathrm{Trace\left( \mathbf {WW}^\dag\right)},\label{eq: SPC}
\end{align}
where $\mathbf {W}= [\mathbf w_1, \mathbf w_2, \dots\mathbf w_G].$
The power radiated by each
antenna element is  a  linear combination of all precoders and reads as
\cite{Yu2007}\begin{align}\label{eq: PAC}
P_n = \left[\sum_{k=1}^G \mathbf w_k \mathbf w_k^\dag \right]_{nn} =\left[ \mathbf {WW}^\dag\right]_{nn},
\end{align}
where $n \in \{1\dots  N_t\}$ is the antenna index.
The fundamental difference between the $\mathrm{SPC}$  of \cite{Karipidis2008} and the proposed $\mathrm{PAC}$ is clear in  \eqref{eq: PAC}, where instead of one,  $N_t$ constraints are realized, each one involving all the precoding vectors.


 \section{Sum Rate Maximization }
 In a multicast scenario,  the performance of all the receivers listening to the same  multicast is dictated by the worst rate in the group. A multigroup multicasting scenario, however,  entails the flexibility  to maximize the total  system rate by providing different service levels amongst groups. The multigroup multicast $\max \mathrm{SR}$ optimization  aims  at maximizing the minimum $\mathrm{SINR}$ only within each group while in parallel maximize the sum of the rates of all groups.
Intuitively, this can be achieved by reducing the power of the users that achieve higher $\mathrm{SINR}$ than the minimum  achieved in the group they belong. Additionally, groups that contain compromised users are turned off and their users driven to service unavailability. Subsequently, power is not consumed in order to mitigate the channel conditions. Any remaining power budget is then reallocated to well conditioned and balanced in term of channel conditions groups.
In \cite{Kaliszan2012}, the $\mathrm{SPC}$ max sum rate problem was solved using a two step heuristic iterative optimization algorithm based on the methods of \cite{Karipidis2008} and \cite{stanczak2009fundamentals}.  Therein,  the  $\mathrm{SPC}$ multicast beamforming problem of \cite{Karipidis2008} is iteratively solved with input $\mathrm{QoS}$ targets defined by the worst user per group of the previous iteration. The derived precoders push all the users of the group closer to the worst user thus saving power. Following that, a power redistribution takes place via the sub-gradient method \cite{stanczak2009fundamentals} towards maximising the total system rate.

\subsection{Per-antenna Power Constrained Optimization }
The present work focuses on the per-antenna power constrained sum rate maximization problem,  formally defined as
  \begin{empheq}[box=\fbox]{align}
\mathcal{SR:}\    \max_{\   \ \{\mathbf w_k \}_{k=1}^{G}}  &\sum_{i=1}^{N_u} \log_2\left(1+\gamma_i \right) & \notag\\
\mbox{subject to: } & \gamma_i  = \min_{m\in G_k}\frac{|\mathbf w_k^\dag \mathbf h_m|^2}{\sum_{l\neq k }^G |\mathbf w_l^\dag\mathbf h_m|^2+\sigma_m^2 }, &\label{const: SR SINR}\\
&\forall i \in\mathcal{G}_k, k,l\in\{1\dots G\},\notag\\
 \text{and to: } & \left[\sum_{k=1}^G  \mathbf w_k\mathbf w_k^\dag  \right]_{nn}  \leq P_n, \label{eq: SR PAC}\\%
 &\forall n\in \{1\dots N_{t}\},\notag
 \end{empheq}
Problem $\mathcal{SR}$ receives as input the per-antenna power constraint vector $\mathbf p_{ant} = [P_1, P_2\dots P_{N_t}]$. Following the common  in the literature notation for ease of reference, the  optimal objective value of $ \mathcal{SR}$ will be denoted as $c^*=\mathcal{SR}(  \mathbf p_{ant})$ and the associated optimal point as $\{\mathbf w_k^\mathcal{SR}\ \}_{k=1}^{G}$. The novelty of the $\mathcal{SR}$  lies in the $\mathrm{PAC}s$, i.e.
\eqref{eq: SR PAC} instead of the conventional SPCs
 proposed in \cite{Kaliszan2012}. To the end of solving this problem, a  heuristic algorithm is proposed. By utilizing recent results \cite{Christopoulos2014}, the new algorithm calculates the per-antenna power constrained precoders.  More specifically, instead of solving the $\mathrm{QoS}$ sum power minimization problem of \cite{Karipidis2008}, the proposed algorithm  calculates the $\mathrm{PAC}$ precoding vectors by solving the per-antenna power minimization  problem \cite{Christopoulos2014}:
\begin{empheq}[box=\fbox]{align}
\mathcal{Q:} \min_{\ r, \ \{\mathbf w_k \}_{k=1}^{G}}  &r& \notag\\
\mbox{subject to } & \frac{|\mathbf w_k^\dag \mathbf h_i|^2}{\sum_{l\neq k }^G | \mathbf w_l^\dag\mathbf h_i|^2 + \sigma^2_i}\geq \gamma_i, \label{const: Q SINR}\\
&\forall i \in\mathcal{G}_k, k,l\in\{1\dots G\},\notag\\
\text{and to}\ \ \ \ \ &\frac{1}{P_n} \left[\sum_{k=1}^G  \mathbf w_k\mathbf w_k^\dag \right]_{nn} \leq  r,\\
& \forall n\in \{1\dots N_{t}\}, \notag
 \end{empheq}
 where $r\in\mathbb{R}^+$. Problem $\mathcal{Q }$ receives as input $\mathrm{SINR}$ the  target  vector $\mathbf g = [\gamma_1,\gamma_2, \dots \gamma_{N_u}]$, that is the individual  $\mathrm{QoS}$ constraints of each user,   as well as the per-antenna power constraint vector $\mathbf p_{ant} .$ Let the  optimal objective value of $ \mathcal{Q}$  be denoted as $r^*=\mathcal{Q}(\mathbf g,  \mathbf p_{ant})$ and the associated optimal point as $\{\mathbf w_k^\mathcal{Q}\ \}_{k=1}^{G}$. This problem is solved using the well established methods of $\mathrm{SDR}$ and Gaussian randomisation \cite{Luo2010}. A more detailed description  of the solution of $ \mathcal{Q}$ can be found in \cite{Christopoulos2014ICC,Christopoulos2014} and is herein omitted for shortness.

Let us rewrite the precoding vectors calculated from $\mathcal{Q }$  as $\{\mathbf w_k^\mathcal{Q} \}_{k=1}^{G}=\{\sqrt{p_k}\mathbf  v_k \}_{k=1}^{G}$ with $ ||v_k ||^2_2=1 $ and $\mathbf p =\left[p_1\dots p_k\right] $. By this normalization, the beamforming problem can be decoupled into two problems. The calculation of the beamforming directions, i.e. the normalized $\{\mathbf  v_k \}_{k=1}^{G}$, and the power allocation over the existing groups, i.e. the calculation of $\mathbf p_k$. Since the exact solution of $\mathcal{SR}$ is not straightforwardly obtained, this decoupling allows for a two step optimization. Under general unicasting assumptions, the $\mathrm{SR}$ maximizing power allocation under fixed beamforming direction is a convex optimization problem \cite{stanczak2009fundamentals}. However, when multigroup multicasting is considered, the cost function   $F_e = \sum_{k=1 }^G \log\left(1+\min_{i \in\mathcal{G}_k}\left\{\mathrm{SINR}_i\right\}\right  )$ is no longer differentiable due to the $\min_{i \in\mathcal{G}_k}$ operation and one has to adhere to sub-gradient solutions\cite{Kaliszan2012}.

 In the present contribution, the calculation of the beamforming directions is based on $\mathcal{Q }$. Following this, the power reallocation is achieved via the sub-gradient method\cite{stanczak2009fundamentals} under specific modifications that are hereafter described.
The proposed algorithm, presented  in Alg. \ref{Alg: MSR PAC}, is an iterative two step  algorithm. In each step of the process, the $\mathrm{QoS}$ targets $\mathbf g$ are calculated as the minimum target per group of the previous iteration, i.e. $\gamma_i = \min_{i \in\mathcal{G}_k}\left\{\mathrm{SINR}_i\right\}, \forall i \in\mathcal{G}_k, k\in\{1\dots G\} $. Therefore,  the new precoders require equal or less power to achieve the  same system sum rate. Any remaining power is then redistributed amongst the groups to the end of maximizing the total system throughput, via the sub-gradient method \cite{stanczak2009fundamentals}.

Focusing of the latter method, let us denote   $\mathbf s =\{ s_{k}\}_{k=1}^G= \{\log  p_{k}\}_{k=1}^G $, as the logarithmic power vector, the sub-gradient search method reads as
  \begin{align}
 \mathbf s{(l+1)}=\prod_{\mathcal P_{a}}\left[\mathbf s{(l)} - \delta(l) \cdot \mathbf{ r}(l)\right], \label{eq: subgrad}
   \end{align}
   where $\prod_{\mathcal P_{a}}[\mathbf x]$ denotes the projection operation of point $\mathbf x \in \mathbb{R}_{G}$ onto the set $\mathcal P_{a}$.The parameters $\delta(l) $  and $\mathbf{ r}(l)$ are the step of the search and the sub-gradient of the  $\mathcal{SR}$ cost function at the point $\mathbf s(l)$, respectively. The analytic calculation of $\mathbf r(l) $ is given in \cite{Kaliszan2012,stanczak2009fundamentals} and is omitted herein for shortness.

In order to account for the more complicated  $\mathrm{PAC}$s, a the following consideration is substantiated.   The projection operation,  i.e. $\prod_{\mathcal P_{a}}[\cdot]$, constrains each iteration of the sub-gradient to the feasibility set of the $\mathcal{SR}$ problem.  The present investigation necessitates the projection over a per-antenna power constrained set rather than a conventional $\mathrm{SPC}$ set proposed in \cite{Kaliszan2012}. Formally, the herein considered set of $\mathrm{PAC}$s  is defined as
\begin{align}
\mathcal P_{a} = \left \{\mathbf p _{}\in\mathbb{R}^+_G| \left[\sum_{k=1}^G  \mathbf v_k \mathrm{diag}(\mathbf p) \mathbf v_k^\dag  \right]_{nn}  \leq P_n  \right\},\label{eq: proj}
\end{align}
where the element of the power vector $\mathbf p_{}=\exp(\mathbf s)$ with $\mathbf{p}, \mathbf{s}\in\mathbb{R}_G$, represent the power allocated to the corresponding group. It should be stressed that this power is inherently different that the power transmitted by each antenna $\mathbf{p}_{ant}\in\mathbb{R}_{N_t}$. The connection between $\mathbf{p}_{ant}$ and $\mathbf{p}$ is given by the normalized beamforming vectors as easily observed in   \eqref{eq: proj}. The per-antenna constrained projection is formally  defined as
\begin{empheq}[box=\fbox]{align}
\mathcal{P:} \min_{\mathbf p_{}}  &||\mathbf p_{} -\mathbf x||^2_2& \notag\\
\mbox{subject to } &  \left[\sum_{k=1}^G  \mathbf v_k \mathrm{diag}(\mathbf p)\mathbf v_k^\dag \right]_{nn} \leq  P_n,\\
& \forall n\in \{1\dots N_{t}\}, \notag
 \end{empheq}
 where $\mathbf p\in \mathbb{R}^G$ and $\mathbf x = \exp\left(\mathbf s{(l)} - \delta(l) \cdot \mathbf{ r}\left(l\right)\right ) $.
Problem $\mathcal{P}$ is a convex optimization problem and can thus be solved to arbitrary accuracy using standard numerical methods \cite{convex_book}.

Subsequently, the solution of \eqref{eq: subgrad} is given as $ \mathbf s (l+1) = \log\left(\mathbf p^*\right ) $, where $\mathbf p^* = \mathcal P\left(\mathbf p_{ant}, \mathbf x\right)$ is the optimal point of convex problem $\mathcal{P}$. To summarize the solution process, the per-antenna power constrained sum rate maximizing algorithm is presented in Alg. \ref{Alg: MSR PAC}.

 \begin{algorithm}
 \SetAlgoLined 
 \KwIn{
(see Tab.\ref{tab: params}) $\{\mathbf w_{k}^{(0)}\}_{k=1}^G = \sqrt{P_{tot}/(G\cdot N_t)}\cdot\mathbf{1}_{N_t}$  }
 \KwOut{   $ \{\mathbf w_{k}^{\mathcal{SR}}\}_{k=1}^G  $}
 \Begin{
 \While{ $\mathcal{SR}$ does not converge}
{$i=i+1;$\\ \textbf{\textit{\uline{Step 1:}}}  Solve  $r^*=\mathcal{Q}(\mathbf g_{(i)} ,  \mathbf p)$ to calculate $\{\mathbf w_{k}^{(i)}\}_{k=1}^G$. The input $\mathrm{SINR}$ targets $\mathbf g_{(i)} $ are given by the minimum $\mathrm{SINR}$ per group, i.e. $\gamma_i = \min_{i \in\mathcal{G}_k}\left\{\mathrm{SINR}_i\right\}, \forall i \in\mathcal{G}_k, k\in\{1\dots G\}$.\\
\textbf{\textit{\uline{Step 2:}}} Initialize the sub-gradient search algorithm as: $\mathbf p ^{(i)} = \{p_{k}\}_{k=1}^G =\{||\mathbf w_{k}^{(i)}||^2_2\}_{k=1}^G $, $ \mathbf s^{(i)} = \{s_{k}^{}\}_{k=1}^G = \{\log  p_{k}\}_{k=1 }^G $, $\{\mathbf v_{k}^{(i)}\}_{k=1}^G = \{\mathbf w_{k}^{(i)}/p_{k}^{(i)}\}_{k=1}^G$.\\
\textbf{\textit{\uline{Step 3:}}} Calculate one iteration of the sub-gradient  power control algorithm \\
 $\mathbf s^{(i+1)}=\prod_{\mathcal P_{a}}\left[\mathbf s^{(i)} - \delta \cdot \mathbf{ r}(i)\right]$  where $\mathbf s = \log(\mathbf p)$, $\mathcal P_{a} = \left \{\mathbf p _{}\in\mathbb{R}^+_G| \left[\sum_{k=1}^G  \mathbf v_k \mathrm{diag}(\mathbf p)  \mathbf v_k^\dag  \right]_{nn}  \leq P_n  \right\} $\\
 \textbf{\textit{\uline{Step 4:}}} Calculate the current throughput:
 $c^{*} = \mathcal{SR} \left(\mathbf p _{ant} \right)$ with
 $\{\mathbf w_{k}^{\mathcal{SR}}\}_{k=1}^G = \{\mathbf w_{k}^{(i+1)}\}_{k=1}^G = \{\mathbf v_{k}^{(i)}\exp(s^{(i+1)}_k)\}_{k=1}^G$
 }
}\label{Alg: MSR PAC}
\caption{  Sum-rate maximizing multigroup multicasting under per-antenna power constraints.}
\end{algorithm}

\begin{table}
\caption{Input Parameters}
\centering
\begin{tabular}{l|l|l}
\textbf{Parameter}  &\textbf{Symbol} &\textbf{Value}\\\hline
 Sub-gradient Iterations  & $l_{max}$ &  $1$   \\
 Sub-gradient step &$\delta$&$0.4$\\
 Gaussian Randomizations&$N_{rand}$&$100$\\
 Total Power at the $T_x$&$P_{tot}$&$[-20:20]$~dBW\\
 Per-antenna constraints &$ \mathbf p_{ant}$ & $P_{tot}/N_t$\\
User Noise variance& $\sigma_i^2$ &$1, \ \forall i \in\{1\dots N_{u}\}$
   \\\hline
\end{tabular}
\label{tab: params}
\end{table}
\subsection{Complexity \& Convergence Analysis }\label{sec: complexity}
 An important discussion involves the complexity of the proposed algorithm. The complexity of the techniques employed  to approximate a solution of  the highly complex, NP-hard  multigroup multicast problem under $\mathrm{PAC}$s is presented in \cite{Christopoulos2014ICC,Christopoulos2014}. Therein, the computational burden for an accurate approximate solution of  the per-antenna power minimization  problem $ \mathcal {Q}$ has been calculated. In summary, the relaxed power minimization is an $\mathrm{SDP}$ instance with $ G $ matrix variables of $ N_t\times N_t $ dimensions and $N_{u}+N_t$ linear constraints. The present work relies on  the CVX tool \cite{convex_book} which calls  numerical  solvers such as SeDuMi to solve  semi-definite programs.  The interior point methods employed to solve this $\mathrm{SDP}$  require at most $\mathcal{O}(\sqrt{GN_t}\log(1/\epsilon)  $ iterations, where $\epsilon $ is the desired numerical accuracy of the solver. Moreover, in each iteration not more than $\mathcal{O}({G^3 N_t^6 +GN_{t}^{3} +N_{u}GN_t^2})$ arithmetic operations will be performed.      The solver used \cite{convex_book} also exploits the specific structure of matrices hence the actual running time is reduced. Next,  a fixed number of iterations of the Gaussian randomization method is performed\cite{Luo2010}.  In each randomization, a linear problem ($\mathrm{LP}$) is solved    with a worst case complexity of  $\mathcal O( G^{3.5}\log(1/\epsilon) ) $ for an  $\epsilon-$optimal solution. The accuracy of the solution increases with the number of randomizations \cite{Karipidis2008,Sidiropoulos2006,Luo2010}.

Focusing on the proposed algorithm, the main complexity burden originates from the solution of a $\mathrm{SDP}$.     The remaining three steps of Alg. \ref{Alg: MSR PAC}
involve a closed form sub-gradient calculation as given in \cite{stanczak2009fundamentals} and the projection operation, which is a  real valued  least square problem under $N_t$ quadratic inequality $\mathrm{PAC}$s. Consequently, the asymptotic complexity of the derived algorithm is polynomial, dominated by the complexity of the $\mathrm{QoS}$ multigroup multicast problem under $\mathrm{PAC}$s.
The convergence of Alg. 1 is guaranteed given that the chosen step size satisfies the  conditions given in \cite{Kaliszan2012,stanczak2009fundamentals}.

\section{Numerical Results}
In the present section, numerical results are presented to quantify the performance gains of the proposed $\mathrm{SR}$ maximization problem under various channel assumptions.    As benchmark, the original $\mathrm{SPC}$ solutions are re-scaled to respect the $\mathrm{PAC}$s, if and only if a constraint is over satisfied. Re-scaling is achieved by multiplying each line of the precoding matrix with the square root of the inverse level of power over satisfaction of the corresponding antenna, i.e.
\begin{align}
\alpha = \sqrt{\max_n\{\mathbf p_{ant}\} / \left[ \mathbf {WW}^\dag\right]_{nn}} \end{align}
    \subsection{Multigroup multicasting over Rayleigh Channels}\label{sec: performance rayleigh}
 The performance of  $\mathcal{SR}$ in terms of $\mathrm {SR}$ is compared to the performance of the solutions of \cite{Kaliszan2012} in a per-antenna constrained transmitter operating over Rayleigh channels in this paragraph. A system with $N_t = 4$ transmit antennas and  $N_u =8  $  users uniformly allocated to $G = 4 $ groups  is assumed, while the channels are generated as Gaussian complex variable instances with unit variance and zero mean. For every channel instance, the  solutions of the  $\mathrm{SPC}$   \cite{Kaliszan2012} and the proposed $\mathrm{PAC}$ $\max \mathrm {SR}$ are evaluated and compared to the weighted  fair solutions of \cite{Christopoulos2014,Christopoulos2014ICC}.  The exact input parameters employed for the algorithmic solution are presented in Tab. \ref{tab: params}. For fair comparison, the total power constraint $P_{tot}$~[Watts] is equally distributed amongst the transmit antennas when $\mathrm{PAC}$s  are considered, hence each antenna can radiate at most $P_{tot}/N_t $~[Watts].
 The results are averaged over one hundred channel realizations, while the noise variance is normalized to one for all receivers.
 The  achievable  $\mathrm{SR}$
 is plotted   in Fig. \ref{fig:  power SR}
 with respect to the total transmit power  $P_{tot}$ in dBW. Clearly, in a practical  $\mathrm{PAC}$ scenario, the proposed optimization problem outperforms existing solutions over the whole $\mathrm{SNR}$ range.
More significantly,   the gains of the derived solution are more apparent in the high power region. In the low power noise limited region, interferences are not the issue and the fair solutions  perform close to the throughput maximizing solution. On the contrary, in the high power regime,  the interference limited fairness solutions saturate in terms of $\mathrm{SR}$ performance. For $P_{tot} = 20$ dBW, the $\max \mathrm{SR}$ solutions attain gains of more than $30\%$ in terms of $\mathrm {SR}$ over the fair approaches. Interestingly, for the same available transmit power,  the  $\mathrm{PAC}$ optimization proposed herein, attains $20\%$ gains over re-scaled to respect the per-antenna constraints $\max \mathrm{SR}$ solutions. Finally, it clearly noted in Fig. \ref{fig:  power SR} that the reported gains increase with respect to the transmit power.
\begin{figure}[h]
\centering
 \includegraphics[width=1\columnwidth]{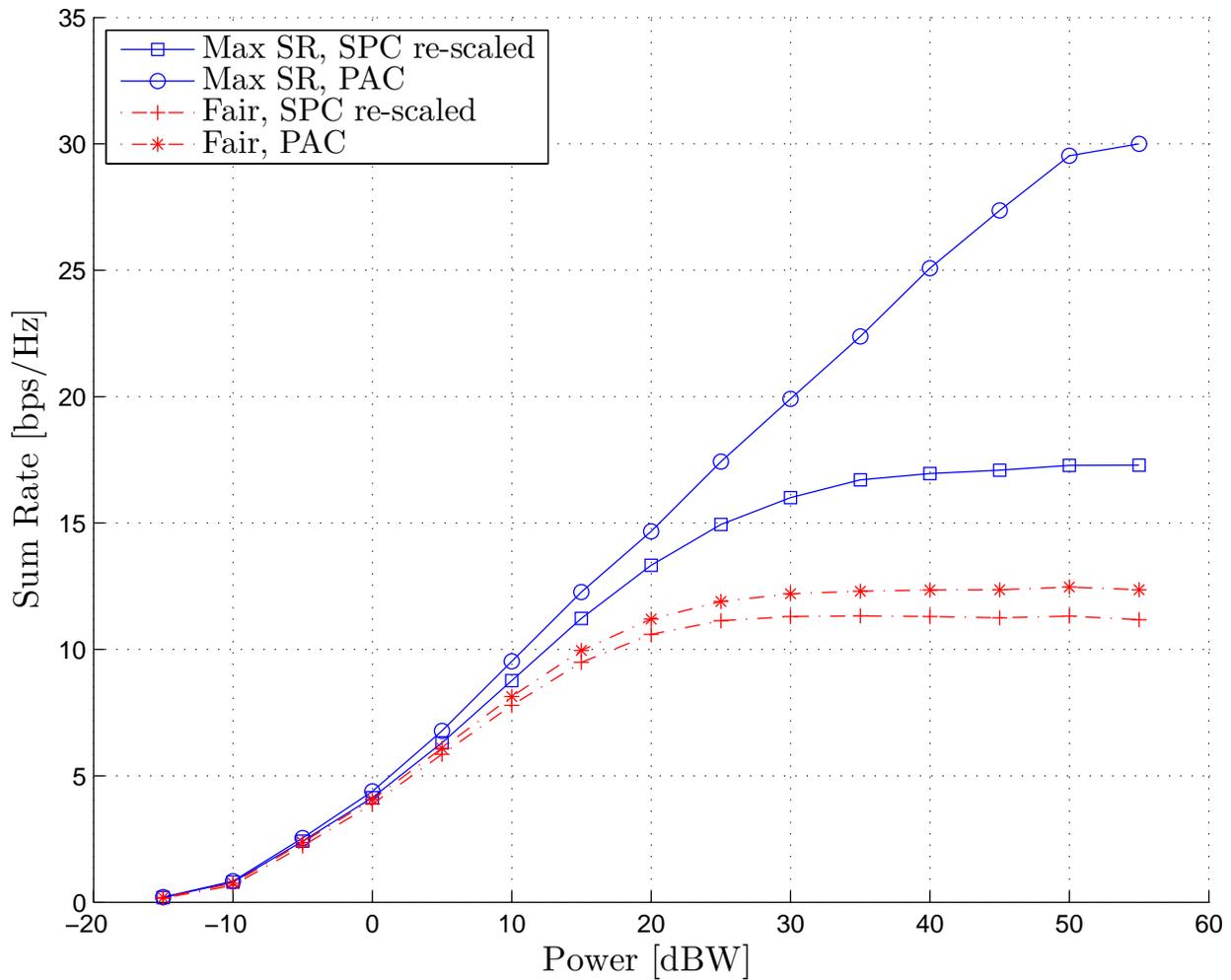}\\
 \caption{System sum rate with $\mathrm{SPC}$ and $\mathrm{PAC}$  versus increasing  total power $P_{tot}$ [dBW].   
}\label{fig:  power SR}
 \end{figure}

A significant issue for the multicast applications is the scaling of the solution  versus an increasing number of receivers per multicast. The increasing number of users per group degrades the performance for the weighted fair problems, as shown in \cite{Christopoulos2014ICC,Christopoulos2014}. For the case tackled herein, the $\max \mathrm{SR}$ solutions are compared to the fairness solutions as depicted in Fig. \ref{fig:  users SR}  with respect to an increasing ratio of users per group $ \rho = N_u / G  $. According to these curves, the $\mathcal{SR}$ solution is exhibiting a higher resilience to the increasing number of users per group, compared to the fair solutions. The re-scaled solutions remain suboptimal in terms of sum rate when compared to proposed solution for any user per group ratio.
\begin{figure}[h]
\centering
\includegraphics[width=1\columnwidth]{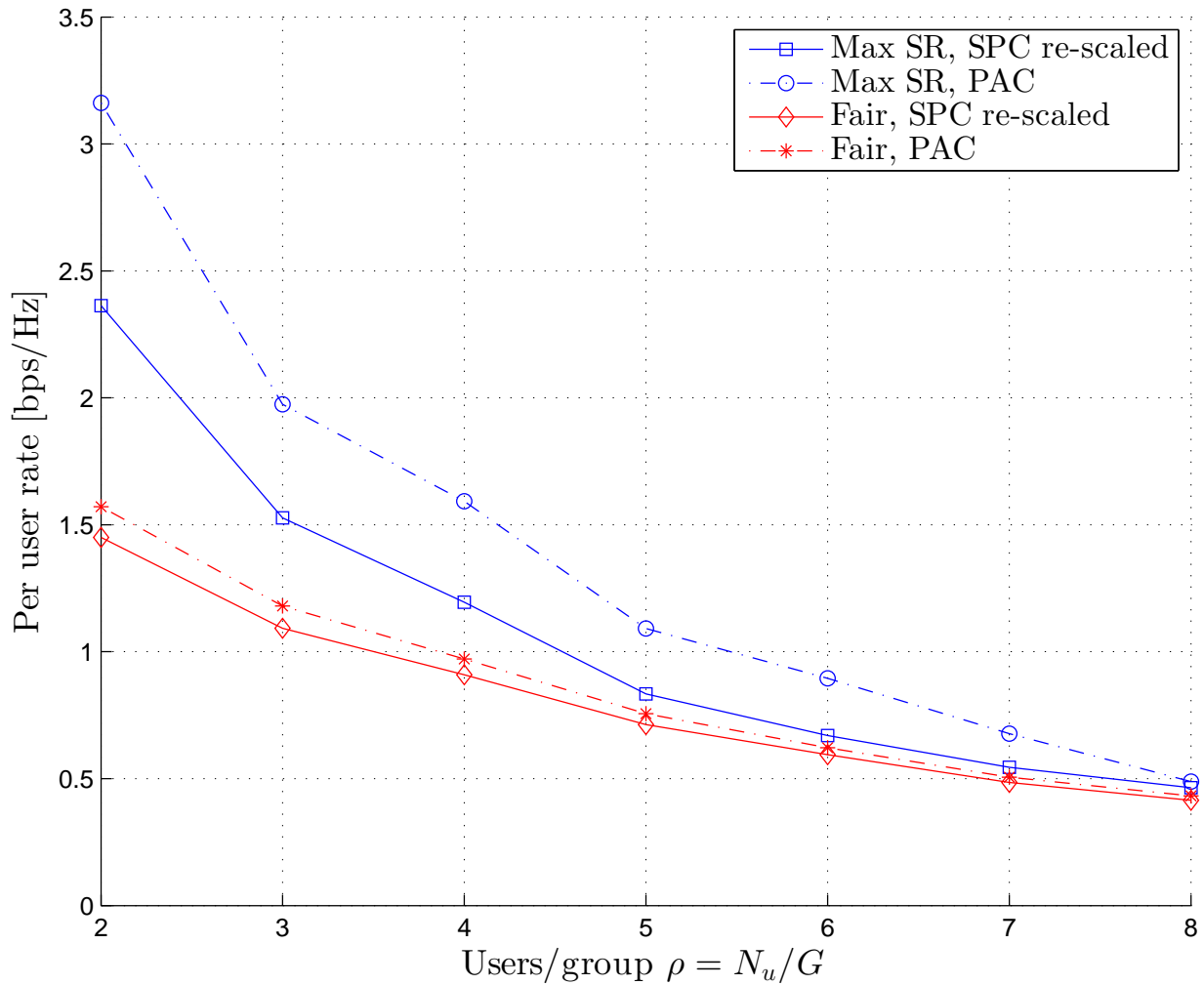}\\
\caption{Sum rate with $\mathrm{SPC}$ and $\mathrm{PAC}$  versus an increasing  ratio of users per group $ \rho = N_u / G  $. 
}\label{fig:  users SR}
\end{figure}
\subsection{Uniform Linear Arrays}
 \begin{figure}[h]
 \centering
 \includegraphics[width=1\columnwidth]{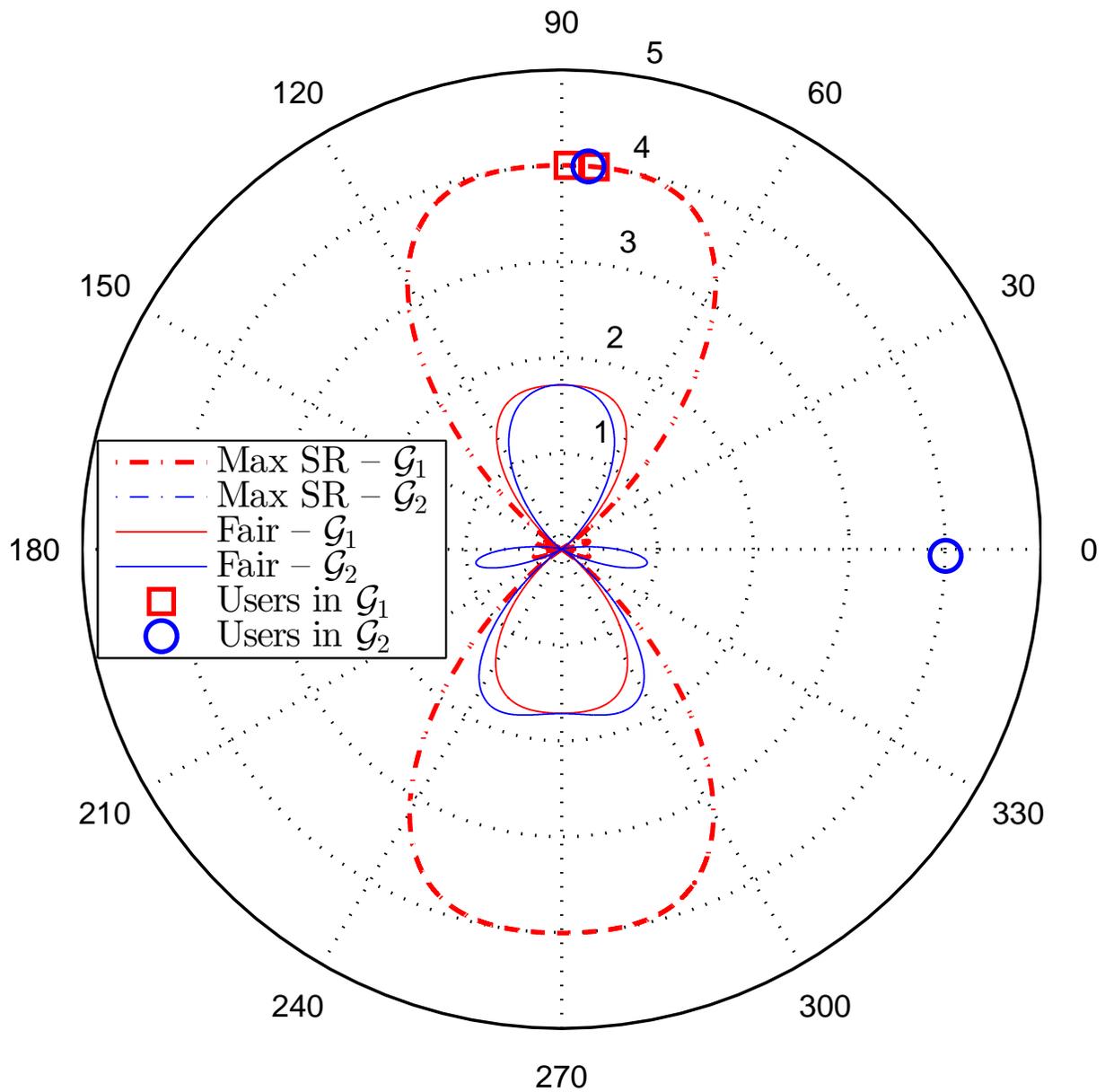}\\
  \caption{User positions  and optimized antenna radiation pattern of $\mathrm{ULA}$ transmitter, under  for  $ \max \mathrm{SR}$ and fairness optimization criteria.}
\label{fig: ULA pos}
\end{figure}
To the end of investigating the sensitivity of the proposed algorithm with respect to the angular separation of co-group users, a uniform linear array ($\mathrm{ULA}$) transmitter is considered. Assuming far-field, line-of-sight conditions, the user  channels  can be  modeled using Vandermonde matrices. Let us consider a $\mathrm{ULA}$ with $N_t = 4$ antennas, serving $4$ users allocated to $2$ distinct groups. The co-group angular separation is $\theta_1 = 5^\circ$ and $\theta_2 = 45^\circ$ for $\mathcal{G}_1$ and $\mathcal{G}_2$ respectively. In Fig. \ref{fig: ULA pos}, the user positions and the optimized radiation pattern for this   is transmitter plotted. The symmetry due to the inherent ambiguity of the $\mathrm{ULA}$ is apparent. Clearly, the fair  beamforming design optimizes the lobes  to provide equal service levels to all users.  The three upper  users (close to the   $ 90^\circ$ angle) receive higher power but also receive  adjacent group interference. The fourth user, despite being in a more favorable in terms of  interference position, is not allocated much power since its performance is constrained by the performance of the almost orthogonal, compromised user.  Remembering that the noise level is equal to one and that the beam pattern is plotted in linear scale, all users achieve a $\mathrm{SINR}$ equal to $0.6$, thus leading to a total $\mathrm{SR}$ of   $1.2$ [bps/Hz]. On the contrary, the $\max \mathrm{SR}$ optimization, shuts down the compromised group (i.e. $\mathcal{G}_2$) and allocates the saved  power to the well conditioned users of $\mathcal{G}_1$.   This way, the system is interference free and each  active user attains a higher service level. The achievable $\mathrm{SNR}$ is equal to $4$ assuming normalized noise, (but only for the two active user of $\mathcal{G}_1$) and leads to a  $\mathrm{SR}$ of more than $4.6$ [bps/Hz]. Consequently,  the  proposed solution attains a $33\%$ of increase in sum rate for the specific scenario, at the expense of sacrificing service availability to the ill conditioned users.

In Fig. \ref{fig: ULA sr}, the performance in terms of the $ \mathrm{SR}$ optimization is investigated versus an increasing angular separation. When co-group users are collocated, i.e. $\theta= 0^\circ$, the highest performance is attained.  As the separation increases, the performance is reduced reaching the minimum when users from different groups are placed in the same position, i.e. $\theta = 45^\circ$. The proposed solution outperforms a re-scaled  to respect the per-antenna constraints, $\mathrm{SPC}$ solution, over the span of the angular separations. Also, the  $\max \mathrm{SR}$ solution performs equivalently to the fair solution under good channel conditions. However, when the angular separation of co-group users increases, the $ \mathrm{SR}$ optimization exploits the  deteriorating  channel conditions and gleans gains of more than $25\%$   over all other solutions.
 \begin{figure}[h]
 \centering
 \includegraphics[width=1\columnwidth]{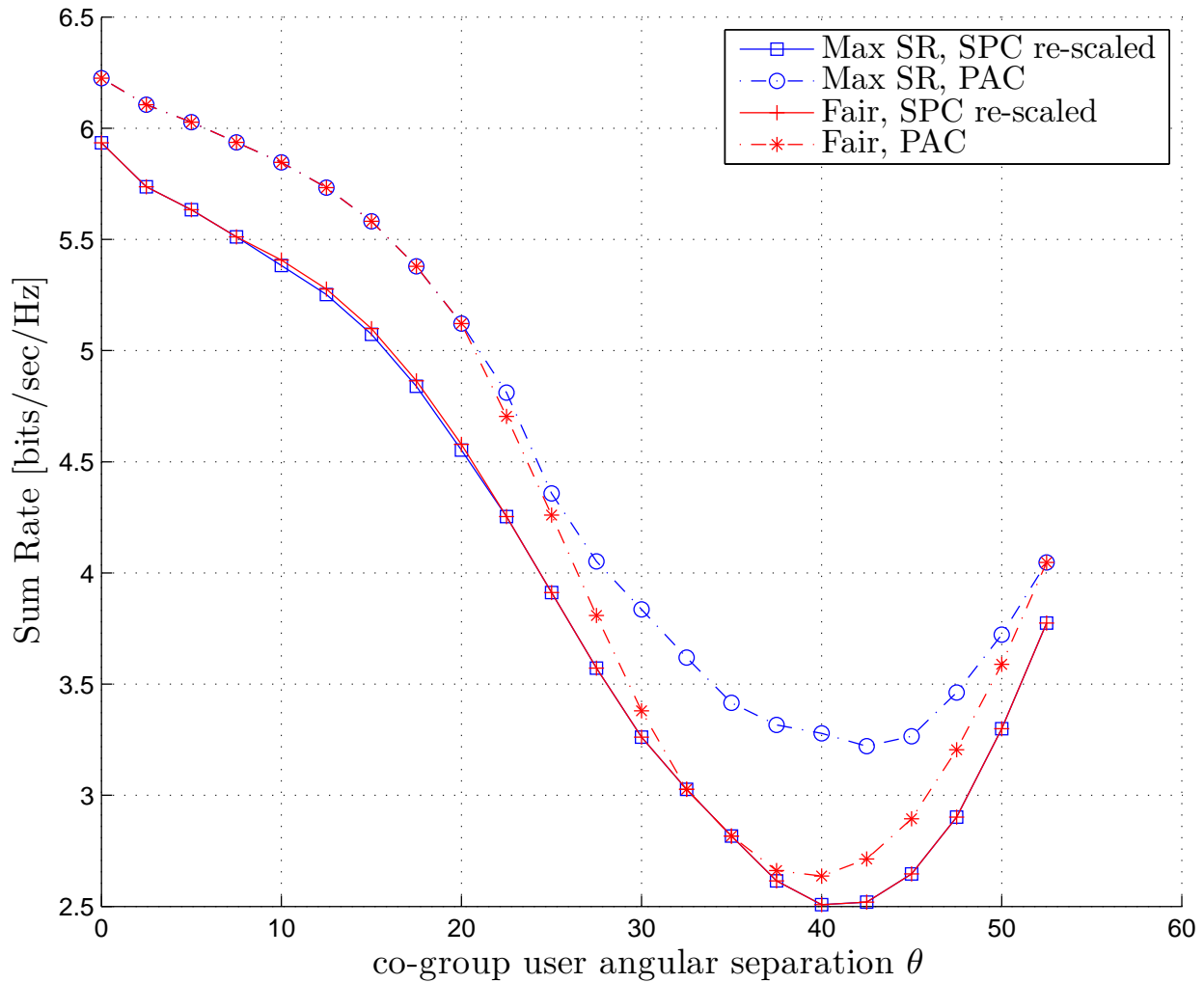}\\
  \caption{Achievable sum rates for  $\mathrm{ULA}$ transmitter with respect to increasing co-group  user angular separation.}
\label{fig: ULA sr}
\end{figure}
\subsection{Sum Rate Maximization Paradigm}

Towards exhibiting the  differences between the weighted fair and the $\max \mathrm{SR}$ designs in the multigroup multicasting context, a small scale paradigm    is presented.
  Let there be a $\mathrm{ULA}$ transmitter that serves eight users   allocated into four groups, as depicted in Fig. \ref{fig: rate bar}.
The attributes of the specific channel instance depict one possible instance of the system where one group, namely $\mathcal{G}_3$, has users with large angular separation while $\mathcal{G}_4$ has  users with similar channels. The rate of each user is plotted in  Fig. \ref{fig: rate bar} for the case of a weighted fair optimization (equal weights are assumed) and for the case of a $\mathrm{SR}$ maximizing optimization. Considering   that each user is constrained by the minimum group rate, the sum rates are given in the legend of the figure. In the weighted fair case,  the common rate at which  all users will receive data is $0.83$ [bps/Hz] leading to a sum rate of $6.64$ [bps/Hz].  The minimum $\mathrm{SINR}$s and hence the minimum rates are balanced between the groups since the  fair optimization considers equal weights. The $\mathrm{SR}$ maximizing optimization, however, reduces  the group that contains the compromised users in order to reallocate this power to the well conditioned group and therefore increase the system throughput  to $9.9$ [bps/Hz]. Consequently, a gain of almost 40\% is realized in terms of total system rate. This gain is  traded-off by driving users in $\mathcal{G}_3$ to the unavailability region.
 \begin{figure}[h]
 \centering
 \includegraphics[width=1\columnwidth]{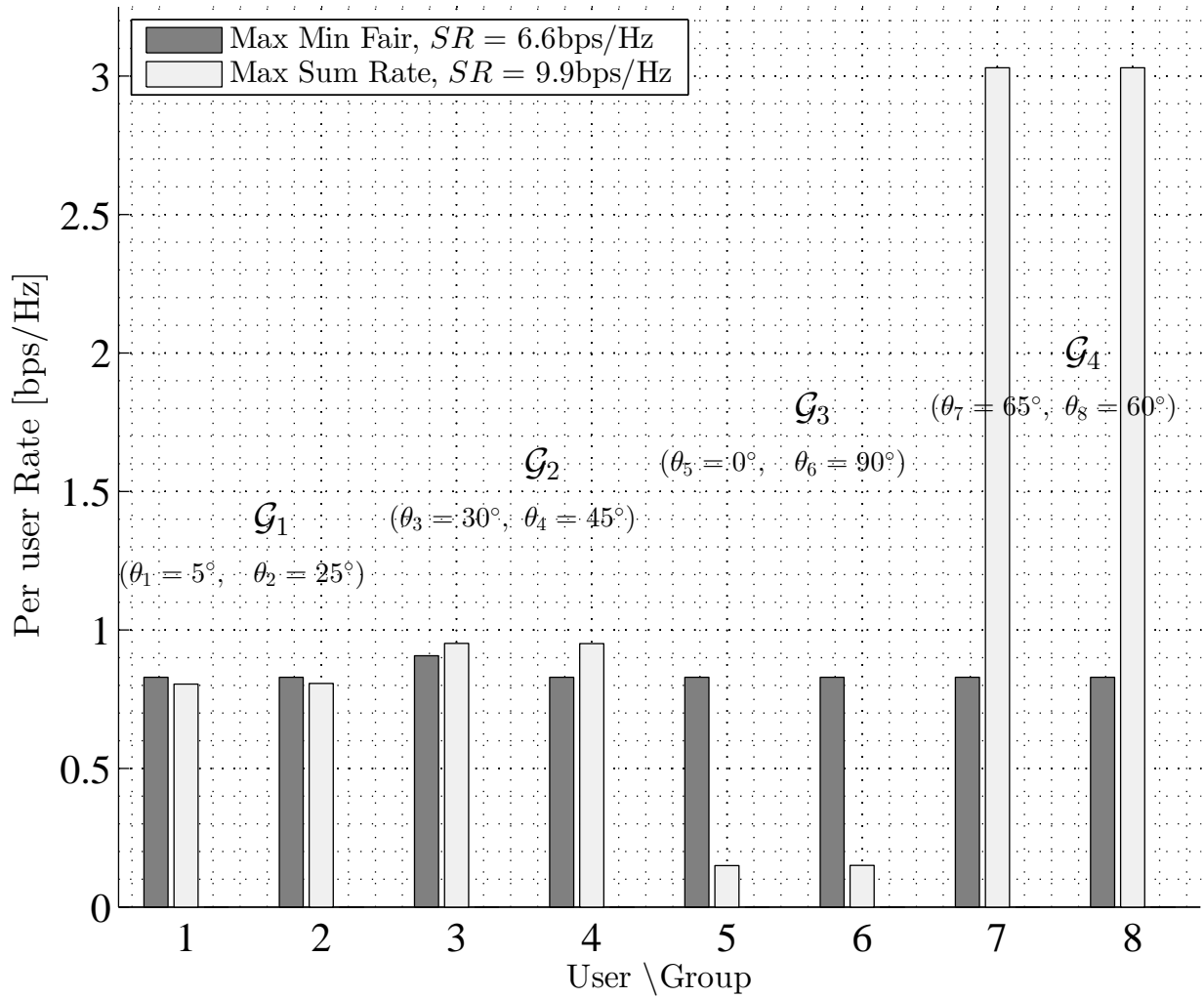}\\
  \caption{Achievable per user rates of  multigroup multicast users under weighted fair and max sum rate optimization. }
\label{fig: rate bar}
\end{figure}
\section{Conclusions \& Future Work}
In the present work, optimum linear precoding vectors are derived when independent sets of common information are transmitted by a per-antenna power constrained array to distinct  co-channel sets of users. In this context, a novel   sum rate maximization  multigroup multicast problem under $\mathrm{PAC}$s  is formulated. A detailed solution for this elaborate  problem is presented based on the well established methods of semidefinite relaxation, Gaussian randomization and sub-gradient power optimization. The performance of the $\mathrm{SR}$ maximizing multigroup multicast optimization is examined under various system parameters and important insights on the system design are gained. Finally, an application paradigm of the new system design is examined.    Consequently, an important practical constraint towards the implementation of throughput maximizing physical layer multigroup multicasting is alleviated. Robust beamforming  as well as availability constrained solutions for multigroup multicast systems are part of future work.
\section*{Acknowledgements}
This work was   partially supported by the National Research Fund, Luxembourg under the projects  ``$\mathrm{CO^{2}SAT}$'' and ``$\mathrm{SEMIGOD}$''.
\bibliographystyle{IEEEtran}
\bibliography{refs/IEEEabrv,refs/conferences,refs/journals,refs/books,refs/references,refs/csi,refs/thesis}

\end{document}